\documentclass[pra,twocolumn]{revtex4}

\usepackage{amsmath}
\usepackage[english]{babel}

\newtheorem{theorem}{Theorem}

\newtheorem{definition}[theorem]{Definition}

\newenvironment{proof}[1][Proof]{\noindent\textbf{#1.} }{\ \rule{0.5em}{0.5em}}

\newtheorem{mytheorem}{Theorem}

\begin{document}

\title{Operator quantum error correction for continuous dynamics}
\author{Ognyan Oreshkov$^{(1,2,3)}$, Daniel A. Lidar$^{(2,3,4,5)}$, Todd A. Brun$%
^{(2,3,5,6)}$} \affiliation{$^{(1)}$Grup de F\'{i}sica Te\`{o}rica,
Universitat Aut\`{o}noma de
Barcelona, 08193 Bellaterra (Barcelona), Spain\\ $^{(2)}$Center for Quantum Information Science \& Technology, $^{(3)}$Department of Physics, $^{(4)}$Department of Chemistry, $^{(5)}$%
Department of Electrical Engineering, $^{(6)}$Communication Science
Institute, \\ University of Southern California, Los Angeles,
California 90089, USA}
\date{\today}

\begin{abstract}
We study the conditions under which a subsystem code is correctable
in the presence of noise that results from continuous dynamics. We
consider the case of Markovian dynamics as well as the general case
of Hamiltonian dynamics of the system and the environment, and
derive necessary and sufficient conditions on the Lindbladian and
system-environment Hamiltonian, respectively. For the case when the
encoded information is correctable during an entire time interval,
the conditions we obtain can be thought of as generalizations of the
previously derived conditions for decoherence-free subsystems to the
case where the subsystem is time dependent. As a special case, we
consider conditions for unitary correctability. In the case of
Hamiltonian evolution, the conditions for unitary correctability
concern only the effect of the Hamiltonian on the system, whereas
the conditions for general correctability concern the entire
system-environment Hamiltonian. We also derive conditions on the
Hamiltonian which depend on the initial state of the environment, as
well as conditions for correctability at only a particular moment of
time. We discuss possible implications of our results for
approximate quantum error correction.
\end{abstract}

\maketitle


\section{Introduction}

Operator quantum error correction (OQEC) \cite{OQEC} is a unified
approach to error correction for noise represented by a completely
positive trace-preserving (CPTP) linear map or noise channel. This
approach uses the most general encoding for the protection of
information---encoding in subsystems \cite{Knill06}. OQEC contains
as special cases the standard quantum error-correction method
\cite{AEC} as well as the methods of decoherence-free subspaces
\cite{DFS} and subsystems \cite{DFs}. Recently, the approach was
generalized to include entanglement-assisted error correction
\cite{EAEC}, resulting in the most general quantum
error-correction formalism presently known for CPTP maps \cite%
{EAOQEC}.

In practice, however, noise is a continuous process and if it can be
represented by a CPTP map, that map is generally a function of time.
Correctability is therefore a time-dependent property. Furthermore,
the evolution of an open system is completely positive if the system
and the environment are initially uncorrelated \cite{Rodriguez:07},
and necessary conditions for CPTP maps are not known. For more
general cases one needs a notion of correctability that can capture
non-CP transformations \cite{LQEC}. Whether completely positive or
not, the noise map is a result of the action of the generator
driving the evolution and possibly of the initial state of the
system and the environment.

Perfect correctability is usually an idealization, since there is
almost always a non-zero probability for uncorrectable errors. For
example, if each qubit in a code undergoes independent errors, no
matter how large the code is, there will always be a non-zero
probability for multi-qubit errors that are not correctable by the
code (although, if this probability per unit time is sufficiently
small, an arbitrarily long information processing task
can be implemented reliably by the use of fault-tolerant techniques \cite{FT}%
). Nevertheless, perfect correctability is a fundamental concept in
the theory of quantum error correction and its understanding is
crucial for the understanding of error correction in realistic
scenarios.

In this paper, we study the question of the conditions under which a
subsystem code is perfectly correctable in the presence of noise
that results from continuous dynamics. We first consider the case
where the subsystem is correctable during an entire time interval
following the encoding, i.e., when the information initially encoded
in the subsystem does not leak out to the environment. Such
conditions are needed in order to understand the mechanisms of
information preservation during continuous processes. If the noise
process is expressed as a CPTP map, the answer is simple---the Kraus
operators have to satisfy the known error-correction conditions at
every moment during the evolution. Our goal is, however, to
understand these conditions in terms of the generator that drives
the evolution---the system-environment Hamiltonian, or in the case
of Markovian evolution the Lindbladian.

We also consider the case where a subsystem can be correctable at a
given moment after the encoding without being correctable during the
entire time interval between the encoding and that moment. This
situation can arise in the case of non-Markovian dynamics, where the
encoded information can flow out to the environment and later return
to the system. We show that the conditions one obtains on the
generator of evolution in this case do not provide non-trivial
information about the properties of the instantaneous dynamics,
except for the global requirement that the linear map resulting from
the dynamics up to the moment in question satisfies the known
error-correction conditions.

Conditions on the generator of evolution have been derived for the
case of decoherence-free subsystems (DFSs) \cite{ShaLid05}, which
are a special type of operator codes. DFSs are \textit{fixed}
subsystems of the system's Hilbert space, inside which all states
evolve unitarily. One generalization of this concept are the
so-called \textit{unitarily correctable subsystems} \cite{OQEC}.
These are subsystems all states inside of which can be corrected via
a unitary operation up to an arbitrary transformation inside the
gauge subsystem. Unlike DFSs, the unitary evolution followed by
states in a unitarily correctable code is not restricted to the
initial subsystem. An even more general concept is that of
\textit{unitarily recoverable} subsystems \cite{OQEC}, for which
states can be recovered by a unitary transformation up to an
expansion of the gauge subsystem. It was shown that
any correctable subsystem is in fact a unitarily recoverable subsystem \cite%
{KS06}. This result reflects the so-called subsystem principle
\cite{Knill06}, according to which protected information is always
contained in a subsystem of the system's Hilbert space. The
connection between DFSs and unitarily recoverable subsystems
suggests that similar conditions on the generators of evolution to
those for DFSs can be derived in the case of general correctable
subsystems. This is the subject of the present paper.

The paper is organized as follows. In Sec.~II, we review the
definitions of correctable subsystems and unitarily recoverable
subsystems. In Sec.~III, we discuss the necessary and sufficient
conditions for such subsystems to exist in the case of CPTP maps. In
Sec.~IV, we derive conditions for the case of Markovian dynamics.
The conditions for general correctability in this case are
essentially the same as those for unitary correctability except that
the dimension of the gauge subsystem is allowed to suddenly
increase. For the case when the evolution is non-correctable, we
conjecture a procedure for tracking the subsystem which contains the
optimal amount of undissipated information and discuss its possible
implications for the problem of optimal error correction. In Sec. V,
we derive conditions on the system-environment Hamiltonian. In this
case, the conditions for continuous unitary correctability concern
only the effect of the Hamiltonian on the system, whereas the
conditions for continuous general correctability concern the entire
system-environment Hamiltonian. In the latter case, the state of the
the noisy subsystem plus environment belongs to a particular
subspace which plays an important role in the conditions. We extend
the conditions to the case where the environment is initialized
inside a particular subspace. In Sec.~VI., we discuss the conditions
under which a subsystem is correctable only at a particular moment
of time. We conclude in Sec.~VII.

\section{Correctable subsystems}

For simplicity, we consider the case where information is stored in only one
subsystem. Then there is a corresponding decomposition of the Hilbert space
of the system,
\begin{equation}
\mathcal{H^{S}}=\mathcal{H^{A}}\otimes \mathcal{H}^{B}\oplus \mathcal{K},
\label{decomposition}
\end{equation}%
where the subsystem $\mathcal{H}^{A}$ is used for encoding of the protected
information. The subsystem $\mathcal{H}^{B}$ is referred to as the gauge
subsystem, and $\mathcal{K}$ denotes the rest of the Hilbert space. In the
formulation of OQEC \cite{OQEC}, the noise process is a CPTP linear map $%
\mathcal{E}:\mathcal{B}(\mathcal{H}^{S})\rightarrow \mathcal{B}(\mathcal{H}%
^{S})$, where $\mathcal{B}(\mathcal{H})$ denotes the set of linear operators
on a finite-dimensional Hilbert space $\mathcal{H}$. Such a map can be
written in the Kraus form \cite{Kraus83}
\begin{equation}
\mathcal{E}(\sigma )=\underset{\alpha }{\sum }M_{\alpha }\sigma M_{\alpha
}^{\dagger },\hspace{0.2cm}\text{for all }\sigma \in \mathcal{B}(\mathcal{H}%
^{S}),  \label{Kraus1}
\end{equation}%
where the Kraus operators $\{M_{\alpha }\}\subseteq \mathcal{B}(\mathcal{H}%
^{S})$ satisfy
\begin{equation}
\underset{\alpha }{\sum }M_{\alpha }^{\dagger }M_{\alpha }=I^{S}.
\label{completeness}
\end{equation}

Let $\mathcal{P}^{AB}(\cdot )$ denote the superoperator projector on $%
\mathcal{B}(\mathcal{H}^{A}\otimes \mathcal{H}^{B})$,
\begin{equation}
\mathcal{P}^{AB}(\cdot )=P^{AB}(\cdot )P^{AB},
\end{equation}%
where $P^{AB}$ is the projector of $\mathcal{H}^{S}$ onto $\mathcal{H}%
^{A}\otimes \mathcal{H}^{B}$,
\begin{equation}
P^{AB}\mathcal{H}^{S}=\mathcal{H}^{A}\otimes \mathcal{H}^{B}.
\end{equation}

We now recall some of the key notions in correctability. The first and
simplest version is one that does not require a recovery (or correction)\
step:

\begin{definition}[Noiseless subsystem]
The subsystem $\mathcal{H}^{A}$ in Eq.~\eqref{decomposition} is called
\textit{noiseless} with respect to the noise process $\mathcal{E}$, if
\begin{gather}
\text{Tr}_{B}\{(\mathcal{P}^{AB}\circ \mathcal{E})(\sigma )\}=\text{Tr}%
_{B}\{\sigma \},  \label{noiselesssystem} \\
\hspace{0.1cm}\text{for all }\sigma \in \mathcal{B}(\mathcal{H}^{S})\text{
such that }\sigma =\mathcal{P}^{AB}(\sigma )\hspace{0.1cm}.  \notag
\end{gather}
\end{definition}

More general is the case when one invokes a correction map to correct the
subsystem:

\begin{definition}[Correctable subsystem]
The subsystem $\mathcal{H}^{A}$ in Eq.~\eqref{decomposition} is called
\textit{correctable} if there exists a correcting CPTP map $\mathcal{R}:%
\mathcal{B}(\mathcal{H}^{S})\rightarrow \mathcal{B}(\mathcal{H}^{S})$, such
that the subsystem is noiseless with respect to the map $\mathcal{R}\circ
\mathcal{E}$:
\begin{gather}
\text{Tr}_{B}\{(\mathcal{P}^{AB}\circ \mathcal{R}\circ \mathcal{E})(\sigma
)\}=\text{Tr}_{B}\{\sigma \},  \label{correctablesystem} \\
\hspace{0.1cm}\text{for all }\sigma \in \mathcal{B}(\mathcal{H}^{S})\text{
such that }\sigma =\mathcal{P}^{AB}(\sigma )\hspace{0.1cm}.  \notag
\end{gather}
\end{definition}

A special case of this is unitary correction:

\begin{definition}[Unitarily correctable subsystem]
The subsystem $\mathcal{H}^{A}$ in Eq.~\eqref{decomposition} is
called \textit{unitarily correctable }when there exists a unitary
correcting map, i.e., when there exists a unitary map
$\mathcal{U}:\mathcal{B}(\mathcal{H}^{S})\rightarrow
\mathcal{B}(\mathcal{H}^{S})$ such that
\begin{gather}
\text{Tr}_{B}\{(\mathcal{P}^{AB}\circ \mathcal{U}\circ \mathcal{E})(\sigma
)\}=\text{Tr}_{B}\{\sigma \},  \label{unitarilycorrectable} \\
\hspace{0.1cm}\text{for all }\sigma \in \mathcal{B}(\mathcal{H}^{S})\text{
such that }\sigma =\mathcal{P}^{AB}(\sigma )\hspace{0.1cm}.  \notag
\end{gather}
\end{definition}

A similar but more general notion is that of a unitarily recoverable
subsystem, for which the unitary $U$ need not bring the erroneous state back
to the original subspace $\mathcal{H}^{A}\otimes \mathcal{H}^{B}$ but can
bring it into a subspace $\mathcal{H}^{A}\otimes \mathcal{H}^{B^{\prime }}$,
with $B$ not necessarily equal to $B^{\prime }$:

\begin{definition}[Unitarily recoverable subsystem]
The subsystem $\mathcal{H}^{A}$ in Eq.~\eqref{decomposition} is
called \textit{unitarily recoverable }when there exists a unitary
map $\mathcal{U}:\mathcal{B}(\mathcal{H}^{S})\rightarrow
\mathcal{B}(\mathcal{H}^{S})$ such that
\begin{gather}
\text{Tr}_{B^{\prime }}\{(\mathcal{P}^{AB^{\prime }}\circ \mathcal{U}\circ
\mathcal{E})(\sigma )\}=\text{Tr}_{B}\{\sigma \},
\label{unitarilyrecoverable} \\
\hspace{0.1cm}\text{for all }\sigma \in \mathcal{B}(\mathcal{H}^{S})\text{
such that }\sigma =\mathcal{P}^{AB}(\sigma )\hspace{0.1cm}.  \notag
\end{gather}
\end{definition}

Obviously, if $\mathcal{H}^A$ is unitarily recoverable, it is also
correctable, since one can always apply a local CPTP map $\mathcal{E}%
^{B^{\prime }\rightarrow B}: \mathcal{B}(\mathcal{H}^{B^{\prime
}})\rightarrow \mathcal{B}(\mathcal{H}^{B})$ which brings all states from $%
\mathcal{H}^{B^{\prime }}$ to $\mathcal{H}^{B}$. (In fact, if the dimension
of $\mathcal{H}^{B^{\prime }}$ is smaller than or equal to that of $\mathcal{H}%
^{B}$, this can always be done by a unitary map, i.e.,
$\mathcal{H}^A$ is unitarily correctable.) In Ref.~\cite{KS06} it
was shown that the reverse is also true---if $\mathcal{H}^A$ is
correctable, it is unitarily recoverable. This equivalence will
provide the basis for our derivation of correctability conditions
for continuous dynamics.

Before we proceed with our discussion, we point out that condition %
\eqref{unitarilyrecoverable} can be equivalently written as \cite{OQEC}
\begin{gather}
\mathcal{U}\circ\mathcal{E}(\rho\otimes\tau)=\rho\otimes\tau^{\prime },%
\hspace{0.4cm} \tau^{\prime }\in\mathcal{B}(\mathcal{H}^{B^{\prime }}),
\label{unitarilyrecoverable2} \\
\hspace{0.2cm} \text{for all }\rho\in \mathcal{B}(\mathcal{H}^A), \hspace{%
0.1cm} \tau\in \mathcal{B}(\mathcal{H}^B)\hspace{0.1cm} .  \notag
\end{gather}

\section{Completely positive linear maps}

An important class of transformations on quantum states consists of
the so-called completely positive (CP) linear maps, also known
simply as quantum operations \cite{NielsenChuang00}. Let
$\mathcal{H}^S$ and $\mathcal{H}^E$
denote the Hilbert spaces of a system and its environment, and let $\mathcal{%
H}=\mathcal{H}^S\otimes\mathcal{H}^E$ be the total Hilbert space. A common
example of a CP map is the transformation that the state of a system
undergoes if the system is initially decoupled from its environment, $%
\rho(0)=\rho^S(0)\otimes \rho^E(0)$, and both the system and environment
evolve according to the Schr\"{o}dinger equation:
\begin{equation}
\frac{d\rho(t)}{dt}=-i[H(t),\rho(t)].  \label{Schrodinger}
\end{equation}
(We work in units in which $\hbar=1$, and assume a generally time-dependent
Hamiltonian.) Equation \eqref{Schrodinger} gives rise to the unitary
transformation
\begin{equation}
\rho(t)=V(t)\rho(0)V^{\dagger}(t),
\end{equation}
with
\begin{equation}
V(t)=\mathcal{T}\text{exp}(-i\int_0^t H(\tau)d\tau),
\end{equation}
where $\mathcal{T}$ denotes time ordering. Under the assumption of
an initially decoupled state of the system and the environment, the
transformation of the state of the system is described by a CPTP map
\begin{equation}
\rho^S(0)\rightarrow \rho^S(t)\equiv\text{Tr}_E\{\rho(t)\}=\underset{\alpha}{%
\sum}M_{\alpha}(t)\rho^S(0)M_{\alpha}^{\dagger}(t),  \notag
\end{equation}
for which the time-dependent Kraus operators $M_{\alpha}(t)\in\mathcal{B}(%
\mathcal{H}^S)$ are given by
\begin{equation}
M_{\alpha}(t)=\sqrt{\lambda_{\nu}}\text{Tr}_E\{I^S\otimes
|\nu\rangle\langle\mu | V(t)\} ,\hspace{0.3cm}\alpha=(\mu,\nu),
\label{Krausoperator}
\end{equation}
where $\{|\mu\rangle \}$ is a basis of the Hilbert space of the environment,
in which the initial environment density operator is diagonal: $\rho^E(0)=%
\underset{\mu}{\sum}\lambda_{\mu}|\mu\rangle\langle \mu|$.

The Kraus representation \eqref{Kraus1} applies to any CP linear
map, which need not necessarily arise from evolution of the type
\eqref{Schrodinger}. This is why, in the following theorem we derive
conditions for discrete CP maps. For correctability under continuous
dynamics, the same conditions must apply at all times, i.e., one can
view the quantities $M_{\alpha }$, $U$, $C_{\alpha }$ and the
subsystem $\mathcal{H}^{B^{\prime }}$ in the theorem as being
implicitly time dependent.

\begin{mytheorem}
The subsystem $\mathcal{H}^A$ in the
decomposition \eqref{decomposition} is
unitarily recoverable
under a CP linear noise process in the form \eqref{Kraus1}, if and
only if there exists a unitary operator
$U\in\mathcal{B}(\mathcal{H}^S)$ such that the Kraus operators
satisfy
\begin{gather}
M_{\alpha}P^{AB}=U^{\dagger}I^{A}\otimes C^{B\rightarrow
B'}_{\alpha},\hspace{0.2cm}C^{B\rightarrow B'}_{\alpha}:
\mathcal{H}^B\rightarrow \mathcal{H}^{B'},\notag\\
\hspace{0.2cm} \forall \alpha. \label{conditionKraus}
\end{gather}
\end{mytheorem}

\begin{proof}
The sufficiency of condition \eqref{conditionKraus} is obvious---using that $%
\rho \otimes \tau $ in Eq.~\eqref{unitarilyrecoverable2} satisfies $\rho
\otimes \tau =P^{AB}\rho \otimes \tau P^{AB}$, it can be immediately
verified that Eq.~\eqref{conditionKraus} implies Eq.~%
\eqref{unitarilyrecoverable2} with $\mathcal{U}=U(\cdot )U^{\dagger }$. Now
assume that $\mathcal{H}^{A}$ is unitarily recoverable and the recovery map
is $\mathcal{U}=U(\cdot )U^{\dagger }$. The map $\mathcal{U}\circ \mathcal{E}
$ in Eq.~\eqref{unitarilyrecoverable2} can then be thought of as having
Kraus operators $UM_{\alpha }$. In particular, condition %
\eqref{unitarilyrecoverable2} has to be satisfied for $\rho =|\psi \rangle
\langle \psi |$, $\tau =|\phi \rangle \langle \phi |$ where $|\psi \rangle
\in \mathcal{H}^{A}$ and $|\phi \rangle \in \mathcal{H}^{B}$ are pure
states.
Notice
that the image of $|\psi \rangle \langle \psi |\otimes |\phi
\rangle \langle \phi |$ under the map $\mathcal{U}\circ \mathcal{E}$ would
be of the form $|\psi \rangle \langle \psi |\otimes \tau ^{\prime }$, only
if all terms in Eq.~\eqref{Kraus1} are of the form
\begin{gather}
UM_{\alpha }|\psi \rangle \langle \psi |\otimes |\phi \rangle \langle \phi
|M_{\alpha }^{\dagger }U^{\dagger }= \label{16}\\
|g_{\alpha }(\psi )|^{2}|\psi \rangle \langle \psi |\otimes |\phi _{\alpha
}^{\prime }(\psi )\rangle \langle \phi _{\alpha }^{\prime }(\psi )|,\hspace{%
0.2cm}g_{\alpha }(\psi )\in C,  \notag
\end{gather}%
where for now we assume that $g_{\alpha }$ and $|\phi _{\alpha
}^{\prime }\rangle $ may depend on $|\psi \rangle $. This follows
from the fact that each of the operators $UM_{\alpha}$ transforms
pure states into pure states, and the (positive) reduced operator on
$\mathcal{H}^A$ of each of the terms \eqref{16} must be proportional
to the same pure state $|\psi\rangle\langle\psi|$ in order for the
total reduced density operator on $\mathcal{H}^A$ to be pure. In
other words,
\begin{equation}
UM_{\alpha }|\psi \rangle |\phi \rangle =g_{\alpha }(\psi )|\psi \rangle
|\phi _{\alpha }^{\prime }(\psi )\rangle ,\hspace{0.2cm}g_{\alpha }(\psi
)\in C,\hspace{0.2cm}\forall \alpha .  \label{edno}
\end{equation}%
But if we impose \eqref{edno} on a linear superposition $|\psi \rangle
=a|\psi _{1}\rangle +b|\psi _{2}\rangle $, ($a,b\neq 0$), we obtain $%
g_{\alpha }(\psi _{1})=g_{\alpha }(\psi _{2})$ and $|\phi _{\alpha
}^{\prime }(\psi _{1})\rangle =|\phi _{\alpha }^{\prime }(\psi
_{2})\rangle $, i.e.,
\begin{equation}
g_{\alpha }(\psi )\equiv g_{\alpha },\hspace{0.2cm}|\phi _{\alpha }^{\prime
}(\psi )\rangle \equiv |\phi _{\alpha }^{\prime }\rangle ,\hspace{0.2cm}%
\forall |\psi \rangle \in \mathcal{H}^{A},\hspace{0.2cm}\forall \alpha .
\end{equation}%
Since Eq.~\eqref{edno} has to be satisfied for all $|\psi \rangle \in
\mathcal{H}^{A}$ and all $|\phi \rangle \in \mathcal{H}^{B}$, we obtain
\begin{gather}
UM_{\alpha }P^{AB}=I^{A}\otimes C_{\alpha }^{B\rightarrow B^{\prime }},%
\hspace{0.2cm}C_{\alpha }^{B\rightarrow B^{\prime }}:\mathcal{H}%
^{B}\rightarrow \mathcal{H}^{B^{\prime }},  \notag \\
\hspace{0.2cm}\forall \alpha .
\end{gather}%
Applying $U^{\dagger }$ from the left yields condition \eqref{conditionKraus}%
.
\end{proof}

We remark that condition \eqref{conditionKraus} is equivalent to the
conditions obtained in Ref.~\cite{OQEC}.

\section{Markovian dynamics}

The most general continuous completely positive time-local evolution of the
state of a quantum system is described by a semi-group master equation in
the Lindblad form \cite{Lin76}:
\begin{eqnarray}
\frac{d\rho(t)}{dt}=-i[H(t),\rho(t)]-\frac{1}{2}\underset{j}{\sum}%
(2L_j(t)\rho(t) L_j^{\dagger}(t)  \notag \\
-L_j^{\dagger}(t)L_j(t)\rho(t)-\rho(t) L_j^{\dagger}(t)L_j(t))\equiv
\mathcal{L}(t)\rho(t).  \label{Lindblad}
\end{eqnarray}
Here $H(t)$ is a system Hamiltonian, $L_j(t)$ are Lindblad operators, and $%
\mathcal{L}(t)$ is the Liouvillian superoperator corresponding to this
dynamics. The general evolution of a state is given by
\begin{equation}
\rho(t_2)=\mathcal{T}\text{exp}(\int_{t_1}^{t_2} \mathcal{L}(\tau)d
\tau)\rho(t_1), \hspace{0.3 cm} t_2>t_1.  \label{evolutionLindblad}
\end{equation}
Such evolution arises from a Hamiltonian interaction with the
environment in the Markovian limit of short bath correlation times
\cite{BrePet02}. The evolution induced by \eqref{Lindblad} is
completely positive and can be thought of as arising from an
infinite sequence of infinitesimal completely positive maps of the
form \eqref{Kraus1}. These maps can depend on time, and therefore
the operators in \eqref{Lindblad} are generally time dependent.

We first derive necessary and sufficient conditions for unitarily
correctable subsystems under the dynamics \eqref{Lindblad}, and then extend
them to the case of unitarily recoverable subsystems.

\subsection{Unitarily correctable subsystems}

In the case of continuous dynamics, the error map $\mathcal{E}$ and the
error-correcting map $\mathcal{U}$ in Eq.~\eqref{unitarilycorrectable} are
generally time dependent. If we set $t=0$ as the initial time at which the
system is prepared, the error map resulting from the dynamics %
\eqref{Lindblad} is
\begin{equation}
\mathcal{E}(t)(\cdot)=\mathcal{T}\text{exp}\left(\int_{0}^{t} \mathcal{L}%
(\tau)d \tau\right)(\cdot).
\end{equation}
Our strategy is now to convert the problem into one of noiseless
subsystems, for which necessary and sufficient conditions have
already been found \cite{ShaLid05}. To this end 
let $\mathcal{U}(t)=U(t)(\cdot)U^{\dagger}(t)$ be the unitary
error-correcting map in Eq.~\eqref{unitarilycorrectable}. We can
define the rotating frame corresponding to $U^{\dagger}(t)$ as the
transformation of each operator as
\begin{equation}
O(t)\rightarrow \widetilde{O}(t)=U(t)O(t)U^{\dagger}(t).
\label{rotatingframe}
\end{equation}
In this frame, the Lindblad equation \eqref{Lindblad} can be written as
\begin{gather}
\frac{d\widetilde{\rho}(t)}{dt}=-i[\widetilde{H}(t)+{H}^{\prime }(t),%
\widetilde{\rho}(t)]-\frac{1}{2}\underset{j}{\sum}(2\widetilde{L}_j(t)%
\widetilde{\rho}(t) \widetilde{L}_j^{\dagger}(t)  \notag \\
-\widetilde{L}_j^{\dagger}(t)\widetilde{L}_j(t)\widetilde{\rho}(t)-%
\widetilde{\rho}(t) \widetilde{L}_j^{\dagger}(t)\widetilde{L}_j(t))\equiv
\widetilde{\mathcal{L}}(t)\widetilde{\rho}(t),  \label{Lindbladrot}
\end{gather}
where $H^{\prime }(t)$ is defined through
\begin{equation}
i\frac{d U(t)}{dt}=H^{\prime }(t) U(t),  \label{defineU}
\end{equation}
i.e.,
\begin{equation}
U(t)=\mathcal{T}\text{exp}\left(-i\int_0^t H^{\prime }(\tau)d\tau\right).
\end{equation}
The CPTP map resulting from the dynamics \eqref{Lindbladrot} is
\begin{equation}
\widetilde{\mathcal{E}}(t)(\cdot)=\mathcal{T}\text{exp}\left(\int_{0}^{t}
\widetilde{\mathcal{L}}(\tau)d \tau\right)(\cdot).
\end{equation}

\begin{mytheorem}
Let $\widetilde{H}(t)$ and
$\widetilde{L}_j(t)$ be the Hamiltonian and the Lindblad operators
in the rotating frame \eqref{rotatingframe} with $U(t)$ given by
Eq.~\eqref{defineU}. Then the subsystem $\mathcal{H}^A$ in the
decomposition \eqref{decomposition} is correctable by $U(t)$
during the evolution \eqref{Lindblad}, if and only if
\begin{gather}
\widetilde{L}_{j}(t)P^{AB}=I^{A}\otimes C^B_j(t),\hspace{0.2cm}\
C^B_j(t)\in\mathcal{B}(\mathcal{H}^B),\hspace{0.2cm}\forall j,
\label{Markov1}
\end{gather}
and
\begin{gather}
\mathcal{P}^{AB}(\widetilde{H}(t)+H'(t))=I^A\otimes
D^B(t),\hspace{0.2cm} D^B(t)\in
\mathcal{B}(\mathcal{H}^B),\label{Markov2}
\end{gather}
and
\begin{gather}
P^{AB}(\widetilde{H}(t)+H'(t)+\frac{i}{2}\underset{j}{\sum}\widetilde{L}_j^{\dagger}(t)\widetilde{L}_j(t))P_{\mathcal{K}}=0,
\label{Markov3}
\end{gather}
for all $t$, where $P_{\mathcal{K}}$ denotes the projector on
$\mathcal{K}$.

\end{mytheorem}

\begin{proof}
Since by definition $U(t)$ is an error-correcting map for subsystem $%
\mathcal{H}^{A}$, if $\mathcal{P}^{AB}(\rho (0))=\rho (0)$, we have $\text{Tr%
}_{B}\{\mathcal{P}^{AB}\circ \widetilde{\mathcal{E}}(\widetilde{\rho }(0))\}=%
\text{Tr}_{B}\{\mathcal{P}^{AB}(\widetilde{\rho }(t))\}=\text{Tr}_{B}\{%
\mathcal{P}^{AB}\circ \mathcal{U}(t)\circ \mathcal{E}(t)(\rho (0))\}=\text{Tr%
}_{B}\{\rho (0)\}=\text{Tr}_{B}\{\tilde{\rho}(0)\}$, i.e, $\mathcal{H}^{A}$
is a noiseless subsystem under the evolution in the rotating frame %
\eqref{Lindbladrot}. Then the theorem follows from
Eq.~\eqref{Lindbladrot} and the conditions for noiseless subsystems
under Markovian dynamics obtained in \cite{ShaLid05}.
\end{proof}

\textbf{Remark.} Conditions \eqref{Markov2} and \eqref{Markov3} can
be used to obtain the operator $H^{\prime }(t)$ (and hence $U(t)$)
if the initial decomposition \eqref{decomposition} is known. Note
that there is a freedom
in the definition of $H^{\prime }(t)$. For example, $D^{B}(t)$ in Eq.~%
\eqref{Markov2} can be any Hermitian operator. In particular, we can choose $%
D^{B}(t)=0$. Also, the term $P_{\mathcal{K}}H^{\prime }(t)P_{\mathcal{K}}$
does not play a role and can be chosen
arbitrarily.
Using that $P_{\mathcal{K}%
}=I-P^{AB}$, we can choose
\begin{gather}
H^{\prime }(t)=-\widetilde{H}(t)-\frac{i}{2}P^{AB}\left( \underset{j}{\sum }%
\widetilde{L}_{j}^{\dagger }(t)\widetilde{L}_{j}(t)\right)   \notag \\
+\frac{i}{2}\left( \underset{j}{\sum }\widetilde{L}_{j}^{\dagger }(t)%
\widetilde{L}_{j}(t)\right) P^{AB},  \label{defineH'}
\end{gather}%
which satisfies Eq.~\eqref{Markov2} and Eq.~\eqref{Markov3}. Using Eq.~%
\eqref{rotatingframe}, Eq.~\eqref{defineU} and Eq.~\eqref{defineH'}, we
obtain the following first-order differential equation for $U(t)$:
\begin{gather}
i\frac{dU(t)}{dt}=-U(t)H(t)-\frac{i}{2}P^{AB}U(t)\left( \underset{j}{\sum }{L%
}_{j}^{\dagger }(t){L}_{j}(t)\right)   \notag \\
+\frac{i}{2}U(t)\left( \underset{j}{\sum }{L}_{j}^{\dagger }(t){L}%
_{j}(t)\right) U^{\dagger }(t)P^{AB}U(t).
\end{gather}%
This equation can be used to solve for $U(t)$ starting from $U(0)=I$.

Notice that, since $\mathcal{H}^{A}$ is unitarily correctable by
$U(t)$, at time $t$ the initially encoded information can be thought
of as contained in the subsystem $\mathcal{H}^{A}(t)$ defined
through
\begin{equation}
\mathcal{H}^{A}(t)\otimes \mathcal{H}^{B}(t)\equiv U^{\dagger }(t)\mathcal{H}%
^{A}\otimes \mathcal{H}^{B},
\end{equation}%
i.e., this subsystem is obtained from $\mathcal{H}^{A}$ in Eq.~%
\eqref{decomposition} via the unitary transformation $U^{\dagger
}(t)$. One
can easily verify that the fact that the right-hand side of Eq.~%
\eqref{Markov1} acts trivially on $\mathcal{H}^{A}$ together with Eq.~%
\eqref{Markov2} are necessary and sufficient conditions for an arbitrary
state encoded in subsystem $\mathcal{H}^{A}(t)$ to undergo trivial dynamics
at time $t$. Therefore, these conditions can be thought of as the conditions
for lack of noise in the instantaneous subsystem that contains the protected
information. On the other hand, the fact that the right-hand side of Eq.~%
\eqref{Markov1} maps states from $\mathcal{H}^{A}\otimes \mathcal{H}^{B}$ to
$\mathcal{H}^{A}\otimes \mathcal{H}^{B}$ together with Eq.~\eqref{Markov3}
are necessary and sufficient conditions for states inside the time-dependent
subspace $U^{\dagger }(t)\mathcal{H}^{A}\otimes \mathcal{H}^{B}$ not to
leave this subspace during the evolution. Thus the conditions of the theorem
can be thought
of
as describing a time-varying noiseless subsystem $%
\mathcal{H}^{A}(t)$.

\subsection{Unitarily recoverable subsystems}

We now extend the above conditions to the case of unitarily
recoverable subsystems. As we pointed out earlier, the difference
between a unitarily \emph{correctable} and a unitarily
\emph{recoverable} subsystem is that in the latter the dimension of
the gauge subsystem may increase. Since the dimension of the gauge
subsystem is an integer, this increase can happen only in a
jump-like fashion at particular moments. Between these moments, the
evolution is unitarily correctable. Therefore, we can state the
following theorem.

\begin{mytheorem}
The subsystem $\mathcal{H}^A$ in
Eq.~\eqref{decomposition} is correctable during the evolution
\eqref{Lindblad}, if and only if there exist times $t_i$, $i=0,1,
2,...$, $t_0=0$, $t_i<t_{i+1}$, such that for each interval
between $t_{i-1}$ and $t_i$ there exists a decomposition
\begin{equation}
\mathcal{H}^S=\mathcal{H}^A\otimes\mathcal{H}^B_i\oplus
\mathcal{K}_i, \hspace{0.4cm}
\mathcal{H}^B_{i}\ni\mathcal{H}^B_{i-1},
\end{equation}
with respect to which the evolution during this interval is
unitarily correctable.
\end{mytheorem}

\textbf{Remark.} An increase of the gauge subsystem at time $t_i$
happens if the operator $C_j(t)$ in Eq.~\eqref{Markov1} obtains
non-zero components that map states from $\mathcal{H}^{B}_i$ to
$\mathcal{H}^{B}_{i+1}$. From that moment on, $t_i\leq t \leq
t_{i+1}$, Eq.~\eqref{Markov1} must hold for
the new decomposition $\mathcal{H}^S=\mathcal{H}^A\otimes\mathcal{H}%
^B_{i+1}\oplus \mathcal{K}_{i+1}$. The unitary $U(t)$ is determined from Eq.~%
\eqref{Markov2} and Eq.~\eqref{Markov3} as described earlier.

The conditions derived in this section provide insights into the
mechanism of information preservation under Markovian dynamics, and
thus could also have implications for the problem of error
correction when perfect recovery is not possible \cite{AQEC1,AQEC}.
For example, it is conceivable that the unitary operation
constructed according to Eq.~\eqref{defineU} with the appropriate
modification for the case of increasing gauge subsystem, may be
useful for error correction also when the conditions of the theorems
are only approximately satisfied. Notice that the generator driving
the effective evolution of the subspace $U^{\dagger }(t)\mathcal{H}^{A}\otimes \mathcal{H}%
^{B}$, whose projector we denote by $P^{AB}(t)\equiv U^{\dagger
}(t)P^{AB}U(t)$, can be written as
\begin{equation}
\mathcal{L}(t)(\cdot )=-i[H_{\text{eff}}(t),\cdot ]+\mathcal{D}(t)(\cdot )+%
\mathcal{S}(t)(\cdot ),
\end{equation}%
where
\begin{gather}
H_{\text{eff}}(t)=H(t)+\frac{i}{2}P^{AB}(t)\left( \underset{j}{\sum }{L}%
_{j}^{\dagger }(t){L}_{j}(t)\right)   \notag \\
-\frac{i}{2}\left( \underset{j}{\sum }{L}_{j}^{\dagger }(t){L}_{j}(t)\right)
P^{AB}(t)
\end{gather}%
is an effective Hamiltonian,
\begin{equation*}
\mathcal{D}(t)(\cdot )=\underset{j}{\sum }L_{j}(t)(\cdot )L_{j}^{\dagger }(t)
\end{equation*}%
is a dissipator, and
\begin{gather}
\mathcal{S}(t)(\cdot )=-\frac{1}{2}P^{AB}(t)\left( \underset{j}{\sum }{L}%
_{j}^{\dagger }(t){L}_{j}(t)\right) P^{AB}(t)(\cdot )  \notag \\
-\frac{1}{2}(\cdot )P^{AB}(t)\left( \underset{j}{\sum }{L}_{j}^{\dagger }(t){%
L}_{j}(t)\right) P^{AB}(t)
\end{gather}%
is a superoperator acting on $\mathcal{B}(U^{\dagger }(t)\mathcal{H}^{AB})$.
The dissipator most generally causes an irreversible loss of the information
contained in the current subspace, which may involve loss of the information
stored in subsystem $\mathcal{H}^{A}(t)$ as well as an increase of the gauge
subsystem. The superoperator $\mathcal{S}(t)(\cdot )$ gives rise to a
transformation solely inside the current subspace. In the case when the
evolution is correctable, this operator acts locally on the gauge subsystem,
but in the general case it may act non-trivially on $\mathcal{H}^{A}(t)$.
The role of the effective Hamiltonian is to rotate the current subspace by
an infinitesimal amount. If one could argue that the information lost under
the action of $\mathcal{D}(t)$ and $\mathcal{S}(t)$ is in principle
irretrievable, then heuristically one could expect that after a single time
step $dt$, the corresponding factor of the infinitesimally rotated (possibly
expanded) subspace will contain the maximal amount of the remaining encoded
information. Note that to keep track of the increase of the gauge subsystem
one would need to determine the operator $C_{j}$ on the right-hand side of
Eq.~\eqref{Markov1} that optimally approximates the left-hand side. Of
course, since the dissipator generally causes leakage of states outside of
the current subspace, the error-correcting map at the end would have to
involve more than just a unitary recovery followed by a CPTP map on the
gauge subsystem. In order to maximize the fidelity \cite{Ore08} of the
encoded information with a perfectly encoded state, one would have to bring
the state of the system fully inside the subspace $\mathcal{H}^{A}\otimes
\mathcal{H}^{B}$. These heuristic arguments, however, require a rigorous
analysis. It is possible that the action of the superoperators $\mathcal{D}%
(t)$ and $\mathcal{S}(t)$ may be partially correctable and thus one may have
to modify the unitary \eqref{defineU} in order to optimally track the
retrievable information. We leave this as a problem for future investigation.

\section{Conditions on the system-environment Hamiltonian}

We now derive conditions for correctability of a subsystem when the
dynamics of the system and the environment is described by the
Schr\"{o}dinger equation \eqref{Schrodinger}. While the CP-map
conditions can account for such dynamics when the states of the
system and the environment are initially uncorrelated, they depend
on the initial state of the environment. Below, we first derive
conditions on the system-environment Hamiltonian that hold for any
state of the environment, and then extend them to the case when the
environment is initialized inside a particular subspace.

We point out that the equivalence between unitary recoverable
subsystems and correctable subsystems has been proven for CPTP maps.
Here, we could have a non-CP evolution since the initial state of
the system and the environment may be entangled. Nevertheless, since
correctability must hold for the case when the initial states of the
system and the environment are uncorrelated, the conditions we
obtain are necessary. They are obviously also sufficient since
unitary recoverability implies correctability.

Let us write the system-environment Hamiltonian as
\begin{equation}
H_{SE}(t)=H_S(t)\otimes I^E + I^S\otimes H_E(t)+H_I(t),
\label{Hamiltonian}
\end{equation}
where $H_S(t)$ and $H_E(t)$ are the system and the environment Hamiltonians
respectively, and
\begin{equation}
H_I(t)=\underset{j}{\sum}S_{j}(t)\otimes E_{j}(t)  \label{HI}
\end{equation}
is the interaction Hamiltonian.

From the point of view of the Hilbert space of the system plus environment,
the decomposition \eqref{decomposition} reads
\begin{gather}
\mathcal{H}=(\mathcal{H}^A\otimes\mathcal{H}^B\oplus\mathcal{K})\otimes
\mathcal{H}^E  \notag \\
=\mathcal{H}^A\otimes\mathcal{H}^B\otimes\mathcal{H}^E\oplus \mathcal{K}%
\otimes\mathcal{H}^E .  \label{decompositionfull}
\end{gather}

\subsection{Conditions independent of the state of the environment}

We will consider again conditions for unitary correctability first, and then
conditions for general correctability.


\subsubsection{Unitary correctability}
In the rotating frame
\eqref{rotatingframe}, the Schr\"{o}dinger equation \eqref{Schrodinger}
becomes
\begin{equation}
\frac{d\widetilde{\rho}(t)}{dt}=-i[\widetilde{H}_{SE}(t)+H^{\prime }(t),%
\widetilde{\rho}(t)].  \label{Schrodingerrot}
\end{equation}

Since in this picture a unitarily correctable subsystem is
noiseless, we can state the following theorem.

\begin{mytheorem}
Consider the evolution \eqref{Schrodinger} driven by the Hamiltonian %
\eqref{Hamiltonian}. Let $\widetilde{H}_{S}(t)$ and $\widetilde{S}_{j}(t)$
be the system Hamiltonian and the interaction operators \eqref{HI} in the
rotating frame \eqref{rotatingframe} with $U(t)$ given by Eq.~\eqref{defineU}%
. Then the subsystem $\mathcal{H}^{A}$ in the decomposition %
\eqref{decomposition} is
unitarily
correctable by $U(t)$ during this evolution, if and
only if
\begin{equation}
\widetilde{S}_{j}(t)P^{AB}=I^{A}\otimes C_{j}^{B}(t),\hspace{0.2cm}\
C_{j}^{B}(t)\in \mathcal{B}(\mathcal{H}^{B}),\hspace{0.2cm}\forall
j, \label{Hamilton1}
\end{equation}%
and
\begin{gather}
(\widetilde{H}_{S}(t)+H^{\prime}(t))P^{AB}=I^{A}\otimes D^{B}(t),\hspace{0.2cm}%
  D^{B}(t)\in \mathcal{B}(\mathcal{H}^{B})
\label{Hamilton2}
\end{gather}
for all t.
\end{mytheorem}

\begin{proof}
With respect to the evolution in the rotating frame \eqref{rotatingframe},
the subsystem $\mathcal{H}^{A}$ is noiseless. The theorem follows from the
conditions for noiseless subsystems under Hamiltonian dynamics \cite%
{ShaLid05} applied to the Hamiltonian in the rotating frame. Note that the
fact that the operator on the right-hand side of Eq.~\eqref{Hamilton2} sends
states from $\mathcal{H}^{A}\otimes \mathcal{H}^{B}$ to $\mathcal{H}%
^{A}\otimes \mathcal{H}^{B}$ implies that the off-diagonal terms of $%
\widetilde{H}_{S}(t)+H^{\prime }(t)$ in the block basis corresponding to the
decomposition \eqref{decomposition} vanish, i.e., $P^{AB}(\widetilde{H}%
_{S}(t)+H^{\prime }(t))P_{\mathcal{K}}=0$.
\end{proof}

\textbf{Remark.} The Hamiltonian $H^{\prime }(t)$ can be obtained
from conditions \eqref{Hamilton1} and \eqref{Hamilton2}. We can
choose $D^{B}(t)=0
$ and define $H^{\prime }(t)=-\widetilde{H}_{S}(t)$, which together with Eq.~%
\eqref{defineU} yields
\begin{equation}
i\frac{dU(t)}{dt}=-U(t)H_{S}(t),
\end{equation}%
i.e.,
\begin{equation}
U^{\dagger }(t)=\mathcal{T}\text{exp}\left( -i\int_{0}^{t}H_{S}(\tau )d\tau
\right) .
\end{equation}%
This simply means that the evolution of the subspace that contains the
encoded information is driven by the system Hamiltonian.

The conditions again can be separated into two parts. The fact that
the right-hand sides of Eq.~\eqref{Hamilton1} and
Eq.~\eqref{Hamilton2} act trivially on $\mathcal{H}^A$ is necessary
and sufficient for the information stored in the instantaneous
subsystem $\mathcal{H}^A(t)$ to undergo trivial dynamics at time
$t$. The fact that the right-hand sides of these equations do not
take states outside of $\mathcal{H}^A\otimes\mathcal{H}^B$ is
necessary and sufficient for states not to leave the subspace $U^{\dagger}(t)%
\mathcal{H}^A\otimes\mathcal{H}^B$ as it evolves.

\subsubsection{Unitary recoverability}
The conditions for
unitary recoverability
are not obtained directly from Theorem 4 in analogy to the case of
Markovian dynamics. Such conditions would certainly be sufficient,
but it turns out that they are not necessary. If after the unitary
recovery operation the dimension of the gauge subsystem
$\mathcal{H}^{B^{\prime }}$ is larger than that of the initial gauge
subsystem $\mathcal{H}^B$, the state of the gauge
subsystem plus environment must belong to a proper subspace of $%
\mathcal{H}^{B^{\prime }}\otimes\mathcal{H}^E$ (because the overall
evolution is unitary and the dimension of the subspace occupied by
the possible states of the system and the environment must be
preserved). Thus it is not necessary that the Hamiltonian acts
trivially on the factor $\mathcal{H}^A$ in
$\mathcal{H}^A\otimes\mathcal{H}^{B^{\prime
}}\otimes\mathcal{H}^E$, but only on the factor $\mathcal{H}^A$ in $\mathcal{%
H}^A\otimes \widetilde{\mathcal{H}}^{BE}$, where $\widetilde{\mathcal{H}}%
^{BE}$ is the proper subspace in question.

\textbf{Example.} To illustrate this point, consider the following
example. Let $\mathcal{H}^S=\mathcal{H}^1\otimes\mathcal{H}^2$ be
the Hilbert space of two qubits with Hilbert spaces $\mathcal{H}^1$
and $\mathcal{H}^2$, respectively. Let the environment consist of a
single qubit, i.e., $\textrm{dim}(\mathcal{H}^E)=2$. We will work in
the rotating frame \eqref{rotatingframe} defined through the
recovering unitary \eqref{defineU} but will drop the tilde for
simplicity of notation and will include the Hamiltonian $H'(t)$ in
the definition of the overall Hamiltonian $H_{SE}(t)$. Let us denote
the basis states of each of the qubit systems by
$|0\rangle^{\sigma}$ and $|1\rangle^{\sigma}$ where the superscript
$\sigma$ labels the qubit ($\sigma=1, 2, E$). Consider the encoding
\eqref{decomposition} with $\mathcal{H}^A=\mathcal{H}^1$ and
$\mathcal{H}^B=\textrm{Span}\{|0\rangle^2\}$. In our basis, the
system-environment Hamiltonian is such that it leaves the state of
qubit $1$ invariant, i.e., its effect on the initial state of the
system plus environment is equivalent to a unitary transformation on
$\mathcal{H}^2\otimes\mathcal{H}^E$. Since the initial state of the
joint system of qubits $2$ and $E$ belongs to the two-dimensional
subspace $\textrm{Span}\{|0\rangle^2 \}\otimes\mathcal{H}^E$ of
$\mathcal{H}^2\otimes\mathcal{H}^E$, the state of these two qubits
at any later time must belong to a two-dimensional subspace of
$\mathcal{H}^2\otimes\mathcal{H}^E$. Let us imagine that the action
of the Hamiltonian up to a given time $t_*$ results in the effective
unitary transformation $I^2\otimes|0\rangle\langle 0|^E+X^2\otimes
|1\rangle\langle 1|^E$ (here $X$ denotes the $\sigma^X$ Pauli
matrix). Then the state of qubits $2$ and $E$ at this moment will
belong to the subspace
$\widetilde{\mathcal{H}}^{BE}=\textrm{Span}\{|0\rangle^2|0\rangle^E,|1\rangle^2|1\rangle^E
\}$. If we assume that there is no restriction on the initial state
of the environment, the reduced density operator on subsystem $2$ at
time $t_*$ can have support on the entire Hilbert space
$\mathcal{H}^2$. In particular, if the environment is initially in
the maximally mixed state $(|0\rangle\langle 0|^E +|1\rangle\langle
1|^E)/2$, the reduced density operator on subsystem $2$ at time
$t_*$ will also be in the maximally mixed state $(|0\rangle\langle
0|^2 +|1\rangle\langle 1|^2)/2$. In other words, the state of qubit
$2$ is not restricted to any particular proper subspace of
$\mathcal{H}^2$ and similarly the state of the environment is not
restricted to any proper subspace of $\mathcal{H}^E$. However, we
cannot argue that the operators $S_j(t_*)$ of the interaction
Hamiltonian \eqref{HI} must have the form $S_j(t_*)=I^1\otimes
C_j^2(t_*)$ (the analogue of Eq.~\eqref{Hamilton1}) or that the
system Hamiltonian in Eq.~\eqref{Hamiltonian} must have the form
$H_S(t_*)=I^1\otimes H^2(t_*)$ (the analogue of
Eq.~\eqref{Hamilton2}). The reason is that since the state of the
entire system plus environment at time $t_*$ belongs to the subspace
$\mathcal{H}^1\otimes\widetilde{\mathcal{H}}^{BE} $, it is necessary
only that $H_{SE}(t_*)$ acts trivially on qubit $1$ when acting on
states in this subspace. For example, if the system Hamiltonian is
$H_S(t_*)=Z^1\otimes Z^2$ (where $Z$ denotes the $\sigma^Z$ Pauli
matrix) and the interaction Hamilonian is
$H_I(t_*)=Z^1\otimes(|11\rangle\langle 11|^{2E}-|00\rangle\langle
00|^{2E})$, qubit $1$ will be effectively acted upon trivially by
the Hamiltonian because the effect of $H_S(t_*)\otimes I^E$ on
states in the subspace $\mathcal{H}^1\otimes
\textrm{Span}\{|0\rangle^2|0\rangle^E,|1\rangle^2|1\rangle^E \}$ is
equivalent to that of $-Z^1\otimes Z^2\otimes I^E$. Note that in the
case of unitary correctability, we can derive necessary conditions
only in terms of the part of the
Hamiltonian that acts on the system because $\mathcal{H}^{B^{\prime }}= \mathcal{H}^{B}$ and $\mathcal{H}%
^{B^{\prime}}\otimes \mathcal{H}^E$ is fully occupied, so the
condition that $\mathcal{H}^A$ in $\mathcal{H}^A\otimes
\mathcal{H}^{B^{\prime}}$ is acted upon trivially must hold for any
state of the environment.

We now proceed to formulate general conditions for
unitary recoverability
under Hamiltonian evolution. Let
\begin{equation}
\mathcal{H}^S=\mathcal{H}^A\otimes \mathcal{H}^{B^{\prime }}\oplus \mathcal{%
K^{\prime }}  \label{maximaldecomposition}
\end{equation}
be a decomposition of the Hilbert space of the system such that the factor $%
\mathcal{H}^{B^{\prime }}\supset \mathcal{H}^{B}$ is such that it
has the largest possible dimension, i.e.,
$\text{dim}(\mathcal{H}^{B^{\prime }})\equiv d^{B^{\prime }}$ is the
largest integer such that
\begin{equation}
\text{dim}(\mathcal{H}^S)=\text{dim}(\mathcal{H}^A)d^{B^{\prime }}+d_{%
\mathcal{K}^{\prime }},
\end{equation}
where $d_{\mathcal{K}^{\prime }}$ is a non-negative integer.

Since the evolution of the state of the system plus the environment is
unitary, at time $t$ the initial subspace $\mathcal{H}^A\otimes\mathcal{H}%
^B\otimes\mathcal{H}^E$ will be transformed to some other subspace
of $\mathcal{H}^S\otimes\mathcal{H}^E$ which is unitarily related to
the initial one. Applying the unitary recovery operation $U(t)$
returns this subspace to the form $\mathcal{H}^A\otimes\widetilde{\mathcal{H}%
}^{BE}(t)$, where $\widetilde{\mathcal{H}}^{BE}(t)$ is a subspace of $%
\mathcal{H}^{B^{\prime }}\otimes\mathcal{H}^E$. Clearly, there exists a
unitary operator $W_0(t): \mathcal{H}^{B^{\prime }}\otimes\mathcal{H}%
^E\rightarrow \mathcal{H}^{B^{\prime }}\otimes\mathcal{H}^E$ that maps this
subspace to the initial subspace $\mathcal{H}^B\otimes \mathcal{H}^E$:
\begin{equation}
W_0(t) \widetilde{P}^{BE}(t) W_0^{\dagger}(t)=P^{BE}.
\label{defineW}
\end{equation}
(Here $\widetilde{P}^{BE}(t)$ denotes the projector on $\widetilde{\mathcal{H%
}}^{BE}(t)$.) Moreover, since the overall unitary that describes the
evolution is a differentiable function of time, if $U(t)$ is chosen
differentiable, $W_0(t)$ can also be chosen differentiable. Note
that as an operator on the entire Hilbert space, this
unitary has the form $W_0(t)\equiv I^A\otimes W_0^{B^{\prime }E}(t)\oplus I_{%
\mathcal{K^{\prime }}}\otimes I^E$.

Let us define the frame
\begin{equation}
\widehat{O}(t)=W(t)O(t)W^{\dagger}(t),  \label{rotatingframe2}
\end{equation}
where
\begin{equation}
i\frac{dW(t)}{dt}=H^{\prime \prime }(t)W(t).  \label{defineW2}
\end{equation}
Then the evolution driven by a Hamiltonian $G(t)$, in this frame will be
driven by $\widehat{G}(t)+H^{\prime \prime }(t)$.

\begin{mytheorem}
Let $\widetilde{O}(t)$ denote the image of an operator $O(t)\in \mathcal{B}(%
\mathcal{H})$ under the transformation \eqref{rotatingframe} with $U(t)\in
\mathcal{B}(\mathcal{H}^{S})$ given by Eq.~\eqref{defineU} ($H^{\prime
}(t)\in \mathcal{B}(\mathcal{H}^{S})$), and let $\widehat{O}(t)$ denote the
image of $O(t)$ under the transformation \eqref{rotatingframe2} with $W(t)$
given by Eq.~\eqref{defineW2}. Let $P^{ABE}$ be the projector on $\mathcal{H}%
^{A}\otimes \mathcal{H}^{B}\otimes \mathcal{H}^{E}$. The subsystem $\mathcal{%
  H}^{A}$ in the decomposition \eqref{decompositionfull} is
unitarily
recoverable by $U(t)$ during the evolution driven by the
system-environment Hamiltonian $H_{SE}(t)$, if and only if there exists $H^{\prime \prime }(t)\in \mathcal{B}%
(\mathcal{H}^{B^{\prime }}\otimes \mathcal{H}^{E})$, where $\mathcal{H}%
^{B^{\prime }}$ was defined in \eqref{maximaldecomposition}, such that
\begin{gather}
(\widehat{\widetilde{H}}_{SE}(t)+\widehat{H}^{\prime }(t)+H^{\prime
\prime}(t))P^{
ABE}=I^{A}\otimes D^{BE}(t),  \label{Hamcond} \\
\hspace{0.2cm}D^{BE}(t)\in \mathcal{B}(\mathcal{H}^{B}\otimes \mathcal{H}%
^{E}),\hspace{0.2cm}\forall t.  \notag
\end{gather}
\end{mytheorem}

\begin{proof}
Assume that the information encoded in $\mathcal{H}^{A}$ is
unitarily recoverable by $U(t)$. Consider the evolution in the frame
defined through the unitary operation $W(t)U(t)$, where
$W(t)=W_0(t)$ for some differentiable $W_0(t)$ that satisfies the
property \eqref{defineW}. In this frame, which can be obtained by
consecutively
applying the transformations \eqref{rotatingframe} and \eqref{rotatingframe2}%
, the Hamiltonian is $\widehat{\widetilde{H}}_{SE}(t)+\widehat{H}^{\prime
}(t)+H^{\prime \prime }(t)$. Under this Hamiltonian, the subsystem $\mathcal{%
H}^{A}$ must be noiseless and no states should leave the subspace $\mathcal{H%
}^{A}\otimes \mathcal{H}^{B}\otimes \mathcal{H}^{E}$. It is straightforward
to see that the first requirement means that $\mathcal{H}^{A}$ must be acted
upon trivially by all terms of the Hamiltonian, hence the factor $I^{A}$ on
the right-hand side of Eq.~\eqref{Hamcond}. At the same time, the subspace $%
\mathcal{\mathcal{H}^{B}\otimes \mathcal{H}^{E}}$ must be preserved by the
action of the Hamiltonian, which implies that the factor $D^{BE}(t)$ on the
right-hand side of Eq.~\eqref{Hamcond} must send states from $\mathcal{H}%
^{B}\otimes \mathcal{H}^{E}$ to $\mathcal{H}^{B}\otimes
\mathcal{H}^{E}$. Note that this implies that the off-diagonal terms
of the Hamiltonian in the block form corresponding to the
decomposition \eqref{decompositionfull} must vanish, i.e.,
$P^{ABE}(\widehat{\widetilde{H}}_{SE}(t)+\widehat{H}^{\prime
}(t)+H^{\prime \prime}(t))P_{\perp}^{ABE}=0$, where $P_{\perp
}^{ABE}$ denotes the projector on $\mathcal{K}\otimes
\mathcal{H}^{E}$. Obviously, these conditions are also sufficient,
since they ensure that in the frame defined by the unitary
transformation $W(t)U(t)$, the evolution of $\mathcal{H}^{A}$
is trivial and states inside the subspace $\mathcal{H}^{B}\otimes \mathcal{H}%
^{E}$ evolve unitarily under the action of the Hamiltonian $D^{BE}(t)$.
Since $W(t)$ acts on $\mathcal{H}^{B^{\prime }}\otimes \mathcal{H}^{E}$,
subsystem $\mathcal{H}^{A}$ is invariant also in the rotating frame %
\eqref{rotatingframe}. This means that $\mathcal{H}^{A}$ is recoverable by
the unitary $U(t)$.
\end{proof}

\textbf{Remark.} Similarly to the previous cases, the unitary
operators $U(t) $ and $W(t)$ can be obtained iteratively from
Eq.~\eqref{Hamcond} if the decomposition \eqref{decomposition} is
given. Since $H^{\prime \prime }(t)$
acts on $\mathcal{H}^{B^{\prime }}\otimes \mathcal{H}^{E}$, from Eq.~%
\eqref{Hamcond} it follows that the operator $\widehat{\widetilde{H}}%
_{SE}(t)+\widehat{H}^{\prime }(t)$ must satisfy
\begin{gather}
(\widehat{\widetilde{H}}_{SE}(t)+\widehat{H}^{\prime}(t))P^{
ABE}=I^{A}\otimes
F^{B^{\prime }E}(t),  \label{determineH'} \\
F^{B^{\prime }E}(t)\in \mathcal{B}(\mathcal{H}^{B^{\prime }}\otimes \mathcal{%
H}^{E}).  \notag
\end{gather}%
At the same time, we can choose $H^{\prime \prime }(t)$ so that $D^{BE}(t)=0$%
. This corresponds to
\begin{equation}
W(t)\widetilde{\mathcal{H}}^{BE}(t)=\mathcal{H}^{B}\otimes \mathcal{H}^{E},
\end{equation}%
where $\widetilde{\mathcal{H}}^{BE}(t)$ was defined in the discussion before
Theorem 5. To ensure $D^{BE}(t)=0$, we can choose
\begin{equation}
H^{\prime \prime }(t)=-\widehat{\widetilde{H}}_{SE}(t)-\widehat{H}^{\prime
}(t)+\mathcal{P}_{\perp }^{ABE}\left( \widehat{\widetilde{H}}_{SE}(t)+%
\widehat{H}^{\prime }(t)\right) ,  \label{determineH''}
\end{equation}%
where $\mathcal{P}_{\perp }^{ABE}(\cdot )=P_{\perp }^{ABE}(\cdot )P_{\perp
}^{ABE}$. For $t=0$ ($U(0)=I$, $W(0)=I$), we can find a solution for $%
\widehat{H}^{\prime }(0)={H}^{\prime }(0)$ from Eq.~\eqref{determineH'},
given the Hamiltonian $\widehat{\widetilde{H}}_{SE}(0)=H_{SE}(0)$. Plugging
the solution in Eq.~\eqref{determineH''}, we can obtain $H^{\prime \prime
}(0)$. For the unitaries after a single time step $dt$ we then have
\begin{equation}
U(dt)=I-iH^{\prime }(0)dt+\mathit{O}(dt^{2}),
\end{equation}%
\begin{equation}
W(dt)=I-iH^{\prime \prime }(0)dt+\mathit{O}(dt^{2}).
\end{equation}%
Using $U(dt)$ and $W(dt)$ we can calculate $\widehat{\widetilde{H}}_{SE}(dt)$
according to Eq.~\eqref{rotatingframe} and Eq.~\eqref{rotatingframe2}. Then
we can solve Eq.~\eqref{determineH'} for $\widehat{H}^{\prime
}(dt)=W(dt)H^{\prime \dagger }(dt)$, which we can use in Eq.~%
\eqref{determineH''} to find $H^{\prime \prime }(dt)$, and so on. Note that
here we cannot specify a simple expression for $\widehat{H}^{\prime }(t)$ in
terms of $\widehat{\widetilde{H}}_{SE}(t)$, since we do not have the freedom
to choose fully $F^{B^{\prime }E}(t)$ in Eq.~\eqref{determineH'} due to the
restriction that $H^{\prime }(t)$ acts locally on $\mathcal{H}^{S}$.

We point out that condition \eqref{Hamcond} again can be understood as
consisting of two parts---the fact that the right-hand side acts trivially
on $\mathcal{H}^A$ is necessary and sufficient for the instantaneous
dynamics undergone by the subsystem $U^{\dagger}(t)W^{\dagger}(t)\mathcal{H}%
^A$ at time $t$ to be trivial, while the fact that it preserves $\mathcal{H}%
^A\otimes\mathcal{H}^{B}\otimes\mathcal{H}^E$ is necessary and sufficient
for states not to leave $U^{\dagger}(t)W^{\dagger}(t) \mathcal{H}^A\otimes%
\mathcal{H}^{B}\otimes\mathcal{H}^E$ as it evolves.

It is tempting to perform an argument similar to the one we
presented for the Markovian case about the possible relation of the
specified recovery unitary operation $U(t)$ and the optimal
error-correcting map in the case of approximate error correction. If
the encoded information is not perfectly preserved, we can construct
the unitary operation $U(t)$ as explained in the comment after
Theorem 5 by optimally approximating Eq.~\eqref{determineH'} and
Eq.~\eqref{determineH''}. However, in this case the evolution is not
irreversible and the information that leaks out of the system may
return to it. Thus we cannot argue that the unitary map specified in
this manner would optimally track the remaining encoded information.

\subsection{Conditions depending on the initial state of the environment}

We can easily extend Theorem 5 to the case when the initial state of the
environment belongs to a particular subspace $\mathcal{H}^{E_0}\in\mathcal{H}%
^E$. The only modification is that instead of $P^{ABE}$ in Eq.~%
\eqref{Hamcond}, we must have $P^{ABE_0}$, where $P^{ABE_0}$ is the
projector on $\mathcal{H}^A\otimes\mathcal{H}^B\otimes \mathcal{H}^{E_0}$,
and on the right-hand side we must have $D^{BE_0}(t)\in \mathcal{B}(\mathcal{H}%
^{B}\otimes\mathcal{H}^{E_0})$.

The following two theorems follow by arguments analogous to those for
Theorem 5. We assume the same definitions as in Theorem 5 [Eq.~%
\eqref{rotatingframe}, Eq.~\eqref{defineU}, Eq.~\eqref{rotatingframe2}, Eq.~%
\eqref{defineW2}], except that in the second theorem we restrict the
definition of $H^{\prime \prime }(t)$.

\begin{mytheorem}
Let $P^{ABE_{0}}$ be the projector on $\mathcal{H}^{A}\otimes \mathcal{H}%
^{B}\otimes \mathcal{H}^{E_{0}}$, where $\mathcal{H}^{E_{0}}\in \mathcal{H}%
^{E}$. The subsystem $\mathcal{H}^{A}$ in the decomposition %
\eqref{decompositionfull} is
unitarily
recoverable by $U(t)\in \mathcal{B}(\mathcal{H}%
^{S})$ during the evolution driven by the system-environment Hamiltonian $%
H_{SE}(t)$ when the state of the environment is initialized inside $\mathcal{%
H}^{E_{0}}$, if and only if there exists $H^{\prime \prime }(t)\in \mathcal{B%
}(\mathcal{H}^{B^{\prime }}\otimes \mathcal{H}^{E})$ such that
\begin{gather}
(\widehat{\widetilde{H}}_{SE}(t)+\widehat{H}^{\prime }(t)+H^{\prime
\prime}(t))P^{
ABE_{0}}=I^{A}\otimes D^{BE_{0}}(t),  \label{Hamconddos} \\
\hspace{0.2cm}D^{BE_{0}}(t)\in \mathcal{B}(\mathcal{H}^{B}\otimes \mathcal{H}%
^{E_{0}}),\hspace{0.2cm}\forall t.  \notag
\end{gather}
\end{mytheorem}

The conditions for unitary correctability in this case require the
additional restriction that $W(t)$ acts on $\mathcal{H}^B\otimes \mathcal{H}%
^E$ and not on $\mathcal{H}^{B^{\prime }}\otimes \mathcal{H}^E$, since in
this case $U(t)$ brings the state inside $\mathcal{H}^A\otimes\mathcal{H}%
^B\otimes\mathcal{H}^E$. Notice that when the state of the environment is
initialized in a particular subspace, we cannot use conditions for unitary
correctability similar to those in Theorem 4. This is because after the
correction $U(t)$, the state of the gauge subsystem plus environment may
belong to a proper subspace of $\mathcal{H}^B\otimes \mathcal{H}^E$ and
tracing out the environment would not yield necessary conditions.

\begin{mytheorem}
Let $P^{ABE_{0}}$ be the projector on $\mathcal{H}^{A}\otimes \mathcal{H}%
^{B}\otimes \mathcal{H}^{E_{0}}$, where $\mathcal{H}^{E_{0}}\in \mathcal{H}%
^{E}$. The subsystem $\mathcal{H}^{A}$ in the decomposition %
\eqref{decompositionfull} is
unitarily
correctable by $U(t)\in B(\mathcal{H}^{S})$
during the evolution driven by the system-environment Hamiltonian $H_{SE}(t)$
when the state of the environment is initialized inside $\mathcal{H}^{E_{0}}$%
, if and only if there exists $H^{\prime \prime }(t)\in \mathcal{B}(\mathcal{%
H}^{B}\otimes \mathcal{H}^{E})$ such that
\begin{gather}
(\widehat{\widetilde{H}}_{SE}(t)+\widehat{H}^{\prime }(t)+H^{\prime
\prime}(t))P^{
ABE_{0}}=I^{A}\otimes D^{BE_{0}}(t),  \label{Hamcondtres} \\
\hspace{0.2cm}D^{BE_{0}}(t)\in \mathcal{B}(\mathcal{H}^{B}\otimes \mathcal{H}%
^{E_{0}}),\hspace{0.2cm}\forall t.  \notag
\end{gather}
\end{mytheorem}

Notice that the conditions of Theorem 6 and Theorem 7 do not depend on the
particular initial state of the environment but only on the subspace to
which it belongs. This can be understood by noticing that different
environment states inside the same subspace give rise to Kraus operators %
\eqref{Krausoperator} which are linear combinations of each other. The
discretization of errors in operator quantum error correction \cite{OQEC}
implies that all such maps will be correctable.

The conditions for correctable dynamics dependent on the state of
the environment could be useful if we are able to prepare the state
of the environment in the necessary subspace. The environment,
however, is generally outside of the experimenter's control.
Nevertheless, it is conceivable that the experimenter may have some
control over the environment (for example, by varying its
temperature), which for certain Hamiltonians could bring the
environment state close to a subspace for which the evolution of the
system is correctable. It is important to point out that according
to a result derived in Ref.~\cite{Ore08}, the error due to imperfect
initialization of the environment will not increase under the
evolution.

\section{Correctability at a particular moment of time}

So far, we have looked at the conditions under which the encoded
information is preserved during an entire time interval following
the encoding. As pointed out earlier, this is not the most general
form of correctability that can occur. It is possible that the
encoded information is lost for a while but is later retrieved. This
clearly cannot happen in the case of Markovian dynamics because
Markovian dynamics is irreversible. However, it can occur in the
case of Hamiltonian dynamics where the information can flow out to
the environment and later return to the system. In this section, we
discuss the conditions on the system-environment Hamiltonian for
this most general type of correctability.

Let the unitary transformation generated by the action of the
system-environment Hamiltonian from time $t=0$ to time $t=T$ be
\begin{equation}
V_{SE}(T)=\mathcal{T}\textrm{exp}(-i\int_0^{T}
H_{SE}(t)dt).\label{VSE}
\end{equation}
The following theorem follows directly from the definition of
unitary recoverability.
\begin{mytheorem}
Let $P^{ABE_0}$ denote the projector on $\mathcal{H}^{A}\otimes \mathcal{H}%
^{B}\otimes \mathcal{H}^{E_{0}}$, where $\mathcal{H}^{E_{0}}\in \mathcal{H}%
^{E}$. Let $\mathcal{H}^{B'}$ be defined as in Eq.~\eqref{maximaldecomposition}. The subsystem $\mathcal{%
  H}^{A}$ in the decomposition \eqref{decompositionfull} is
unitarily recoverable by $U=U^S\otimes I^E$ at time $t=T$ under the
evolution driven by the system-environment Hamiltonian $H_{SE}(t)$
when the state of the environment is initialized inside
$\mathcal{H}^{E_0}$, if and only if
\begin{gather}
UV_{SE}(T)P^{ABE_0}=I^A\otimes C^{BE_0\rightarrow
B'E},\label{theorem8}
\\
C^{BE_0\rightarrow B'E}:
\mathcal{H}^B\otimes\mathcal{H}^{E_0}\rightarrow
\mathcal{H}^{B'}\otimes\mathcal{H}^E.\notag
\end{gather}
\end{mytheorem}

The theorems for the case of unitary correctability or
correctability independent of the state of the environment can be
obtained from Theorem 8 by substituting
$\mathcal{H}^{B'}=\mathcal{H}^B$ and
$\mathcal{H}^{E_0}=\mathcal{H}^E$, respectively.

Theorem 8 can be regarded as a generalization of Theorem 1, which
concerns the process that leads to a particular transformation at a
given moment, rather than the transformation itself. More
specifically, condition \eqref{theorem8} is equivalent to the
condition that all possible CPTP maps obtained through
Eq.~\eqref{Krausoperator}, for the different possible initial
density matrices of the environment with support on
$\mathcal{H}^{E_0}$, satisfy Eq.~\eqref{conditionKraus}. This
equivalence can be obtained by sandwiching both sides of
Eq.~\eqref{theorem8} between all pairs of vectors $|\mu\rangle$ and
$|\nu\rangle$ from an orthonormal basis which spans
$\mathcal{H}^{E}$ and a subset of which spans $\mathcal{H}^{E_0}$.

Note that Theorem 8 imposes conditions on the Hamiltonian
$H_{SE}(t)$ only indirectly---through a condition on the resulting
unitary \eqref{VSE}. At first sight this may not seem too different
from the situation we had before for the case of continuous
correctability, because the conditions in that case (e.g., Theorem
6) could be regarded as equivalent to the requirement that Theorem 8
holds at every moment of time. But precisely because in that case
condition \eqref{theorem8} was imposed for all times, we obtained
non-trivial conditions on the Hamiltonian for all times. Those
non-trivial conditions ensured that, at every moment of time, the
Hamiltonian does not take the information of interest outside the
system.

In this case, the only restriction on the resulting unitary is the
global requirement that at time $T$ the unitary $V_{SE}(T)$
satisfies Eq.~\eqref{theorem8}. But up to any moment $t_0$,
$0<t_0<T$, the unitary $V_{SE}(t)$ can be completely arbitrary
because it can always become of the form that satisfies Theorem 8
during the interval between $t_0$ and $T$. Therefore, if we write
conditions on the Hamiltonian similar to those for continuous
correctability, up to any moment $t_0<T$ these conditions will be
trivial. The only non-trivial condition has a global character and
it is expressed through the condition on $V_{SE}(T)$ as given by
Theorem 8.

\section{Conclusion}

We have derived conditions for correctability of subsystems under
continuous dynamics. We first presented conditions for the case when
the evolution can be described by a CPTP linear map. These
conditions are equivalent to those known for operator codes
\cite{OQEC} except that we consider them for time-dependent noise
processes. We then derived conditions for continuous correctability
for the case of Markovian dynamics and general Hamiltonian dynamics
of the system and the environment. We derived conditions for both
unitary correctability and general correctability, using the fact
that correctable subsystems are unitarily recoverable \cite{KS06}.
For the case of Hamiltonian dynamics, we also considered conditions
for correctability at only a particular moment of time.

The conditions for continuous correctability under both Markovian
and Hamiltonian evolution can be understood as consisting of two
parts---the first is necessary and sufficient for
the absence
of noise inside the instantaneous subsystem that contains the
information, and the second is necessary and sufficient for states
not to leave the subsystem as it evolves with time. In this sense,
the new conditions can be thought of as generalizations of the
conditions for noiseless subsystems to the case where the subsystem
is time dependent.

In the Hamiltonian case, the conditions for continuous unitary
correctability concern only the action of the Hamiltonian on the
system, whereas the conditions for continuous
unitary recoverability
concern the entire system-environment Hamiltonian. The reason for
this is that the state of the gauge subsystem plus the environment
generally belongs to a particular subspace, which does not factor
into a sector belonging to the system and a sector equal to the
Hilbert space of the environment.

We also derived conditions in the Hamiltonian case that depend on the
initial state of the environment. These conditions could be useful, in
principle, since errors due to imperfect initialization of the environment
do not increase under the evolution. Furthermore, these conditions could
provide a better understanding of correctability under CPTP maps, since a
CPTP map that results from Hamiltonian evolution depends on both the
Hamiltonian and the initial state of the environment.

Finally, we discussed the conditions for correctability at only a
particular moment of time. This most general form of correctability
can occur in the case of non-Markovian dynamics where the
information can flow out to the environment but later return to the
system. We showed that the conditions on the generator of evolution
in this case amount to a condition on the overall transformation and
do not provide non-trivial information about the time-local
properties of the dynamics.

We also discussed possible implications of the conditions for
continuous correctability for the problem of optimal recovery in the
case of imperfectly preserved information. We hope that the results
obtained in this paper will provide insight into the mechanisms of
information flow under decoherence that could be useful in the area
of approximate error correction as well.

\section*{Acknowledgments}

OO\ acknowledges support under NSF Grant No. CCF-0524822 and Spanish
MICINN (Consolider-Ingenio QOIT); DAL acknowledges support under NSF
Grant No. CCF-0523675 and NSF grant CCF-0726439; TAB acknowledges
support under NSF Grant No. CCF-0448658.

\end{document}